\documentclass[aps,prl,twocolumn,showpacs,superscriptaddress,amssymb]{revtex4-1}

\usepackage{graphicx}
\usepackage{amsmath}
\usepackage{bm}
\usepackage{color}

\begin{document}

\title{Shear flow in a three-dimensional complex plasma in microgravity conditions}

\author{V. Nosenko}
\email{V.Nosenko@dlr.de}
\affiliation{Institut f\"{u}r Materialphysik im Weltraum, Deutsches Zentrum f\"{u}r Luft- und Raumfahrt (DLR), D-82234 We{\ss}ling, Germany}

\author{M. Pustylnik}
\affiliation{Institut f\"{u}r Materialphysik im Weltraum, Deutsches Zentrum f\"{u}r Luft- und Raumfahrt (DLR), D-82234 We{\ss}ling, Germany}

\author{M. Rubin-Zuzic}
\affiliation{Institut f\"{u}r Materialphysik im Weltraum, Deutsches Zentrum f\"{u}r Luft- und Raumfahrt (DLR), D-82234 We{\ss}ling, Germany}

\author{A. M. Lipaev}
\affiliation{Institute for High Temperatures, Russian Academy of Sciences, Izhorskaya 13/19, 125412 Moscow, Russia}
\affiliation{Moscow Institute of Physics and Technology, Institutsky lane 9, Dolgoprudny, Moscow region, 141700 Russia}

\author{A. V. Zobnin}
\affiliation{Institute for High Temperatures, Russian Academy of Sciences, Izhorskaya 13/19, 125412 Moscow, Russia}

\author{A. D. Usachev}
\affiliation{Institute for High Temperatures, Russian Academy of Sciences, Izhorskaya 13/19, 125412 Moscow, Russia}

\author{H. M. Thomas}
\affiliation{Institut f\"{u}r Materialphysik im Weltraum, Deutsches Zentrum f\"{u}r Luft- und Raumfahrt (DLR), D-82234 We{\ss}ling, Germany}

\author{M. H. Thoma}
\affiliation{I. Physikalisches Institut, Justus-Liebig-Universit\"{a}t Gießen, Heinrich-Buff-Ring 16, 35392 Gießen, Germany}

\author{V. E. Fortov}
\affiliation{Institute for High Temperatures, Russian Academy of Sciences, Izhorskaya 13/19, 125412 Moscow, Russia}

\author{O. Kononenko}
\affiliation{Gagarin Research and Test Cosmonaut Training Center, 141160 Star City, Moscow Region, Russia}

\author{A. Ovchinin}
\affiliation{Gagarin Research and Test Cosmonaut Training Center, 141160 Star City, Moscow Region, Russia}

\date{\today}
\begin{abstract}
Shear flow in a three-dimensional complex plasma was experimentally studied in microgravity conditions using Plasmakristall-4 (PK-4) instrument on board the International Space Station (ISS). The shear flow was created in an extended suspension of microparticles by applying the radiation pressure force of the manipulation-laser beam. Individual particle trajectories in the flow were analysed and from these, using the Navier-Stokes equation, an upper estimate of the complex plasma's kinematic viscosity was calculated in the range of $0.2$--$6.7~{\rm mm^2/s}$. This estimate is much lower than previously reported in ground-based experiments with 3D complex plasmas. Possible reasons of this difference are discussed.
\end{abstract}

\pacs{
52.27.Lw 
}

\maketitle

{\it Introduction.} Shear flows in liquids are ubiquitous in nature and in laboratory experiments, they are important in fundamental science and numerous applications. Shear viscosity is an important characteristic of a liquid which quantifies its resistance to flow; it has a central role in understanding and describing shear flows.

Complex plasmas are suspensions of nanometer to micrometer sized solid particles in a regular plasma \cite{Ivlev_book}. The particles charge up (usually negatively) by collecting electrons and ions from the plasma and interact with each other via a screened Coulomb pair potential. Complex plasmas exist in two-dimensional (2D) and three-dimensional (3D) configurations. They are excellent model systems, which allow studying various phenomena including shear flows, at the most fundamental level of individual particles and in real time \cite{Nosenko:2004,Gavrikov:2005}. Since gravity plays an important role in the balance of forces acting on the particles, microgravity conditions are necessary to obtain large unstrained 3D suspensions of particles. Such conditions are achieved in parabolic flights of specialized aircraft, sounding rockets, and in microgravity laboratories on board the ISS.

Only a few experiments with shear flows in 3D complex plasmas, all of them ground-based, have been reported in the literature \cite{Gavrikov:2005,Vorona:2007,Ivlev:2007}. Reported values of experimentally measured kinematic viscosity of complex plasmas vary in a wide range of $0.8$--$300~{\rm mm^2/s}$, depending on the experimental conditions. Shear flows in 3D complex plasmas in microgravity conditions have not been studied so far.

It is instructive to compare the experimentally measured shear viscosity of complex plasmas with that obtained in molecular dynamics (MD) simulations of Yukawa liquids \cite{Murillo:01,Salin:02,Salin:03,Hamaguchi:02}. However, care should be taken when doing such a comparison. First, the actual particle pairwise interaction potential in a complex plasma is more complicated than the Yukawa potential used in the simulations \cite{Kompaneets_PhD}. Second, simulations use various equilibrium or nonequilibrium methods, whereas experiments with shear flows are by design nonequilibrium.

In this paper, we experimentally study shear flow in a 3D complex plasma in microgravity conditions using the Plasmakristall-4 (PK-4) instrument \cite{Pustylnik:2017} on board the International Space Station (ISS). An (upper) estimate of the shear viscosity of the complex plasma liquid is given and compared with previous experimental results and MD simulations.

{\it Experimental method.} PK-4 is the latest generation of ISS instruments intended to study complex plasmas in microgravity conditions. Compared to its predecessors, it is particularly well suited for studying flow phenomena in liquid 3D complex plasmas. Neon or argon plasma is produced by a direct current (DC) discharge in a long $3$-cm-diameter glass tube. Melamine formaldehyde (MF) microspheres with diameters in the range of $1.3$--$10.4~\mu$m are injected in the plasma from one of the six available dispensers. A particle cloud is then trapped in the middle of the tube by switching the discharge polarity at a frequency of up to $1$~kHz. The particles are illuminated by a thin laser sheet and imaged by two video cameras with slightly overlapping fields of view, which can be combined into one. The instrument and its operation are described in detail in Ref.~\cite{Pustylnik:2017}.

\begin{figure}[tb]
\centering
\includegraphics[width=0.95\columnwidth]{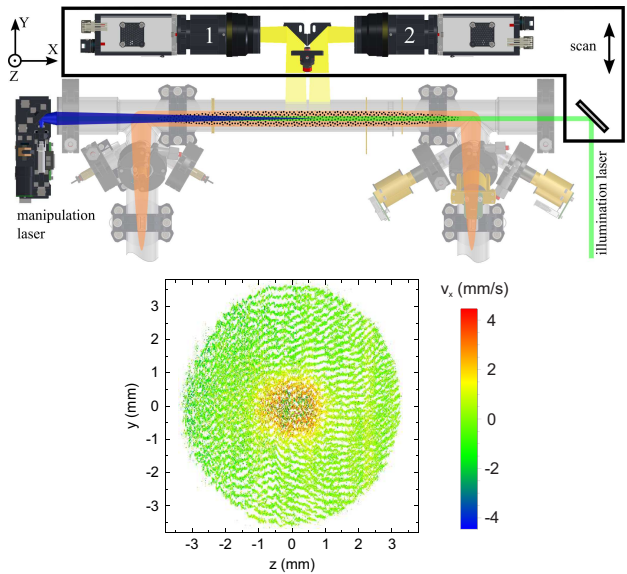}
\caption {\label {vx_y_z} (top) Schematic diagram of the experimental setup. Shear flow in the particle suspension is induced by the manipulation-laser beam. The particle observation cameras 1 and 2 are labeled accordingly. (This image was reproduced with modification from \cite{Pustylnik:2017}, with the permission of AIP Publishing.) (bottom) Profile of the particle longitudinal velocity $v_x(y,z)$ reconstructed from the suspension scan (in the $y$ direction). The higher-velocity area in the middle roughly corresponds to the cross section of the manipulation-laser beam.
}
\end{figure}

The experiments reported in this paper were performed after the PK-4 hardware had been upgraded by installing the so-called Experimental Interface (details will be published elsewhere), which allows experimental regimes without residual gas flow. This helped to minimize undesired disturbances of the particle suspension. The experimental setup is shown schematically in Fig.~\ref{vx_y_z}. The main experimental parameters are listed in Table~\ref{tab1}. We used Ne plasma; the gas pressure was in the range of $15$--$60$~Pa. The DC discharge current was $0.5$~mA and the maximum DC voltage was $1.5$~kV. The electron density $n_e$ and temperature $T_e$ were estimated on the tube axis in the middle of the working area, see Ref.~\cite{Pustylnik:2017} for more details. We used MF microparticles with diameters of $3.38\pm0.07$ and $6.86\pm0.12~\mu$m. They were trapped using polarity switching with a frequency of $500$~Hz and duty cycle $\simeq50$\%. For the particle charge $Q$, we adopted the values reported for our experimental conditions in Ref.~\cite{Antonova:2019}. For the particle neutral gas drag rate $\gamma$, we used the Epstein expression \cite{Epstein:1924}. The Wigner-Seitz radius of the particle suspension was calculated as $a=r_0/1.79$, where $r_0$ is the first peak position of the pair correlation function $g(r)$ measured in 2D cross sections of the particle suspension \cite{Liu:2015}. The particle number density was calculated as $n=3(4\pi a^3)^{-1}$. The coupling parameter was estimated as $\Gamma=Q^2(4\pi\epsilon_0aE_k)^{-1}$, where $E_k$ is the mean kinetic energy of the random motion of particles (on top of the mean flow velocity). In some experimental runs, the particle suspension was scanned by synchronously moving the illumination laser sheet and video cameras across the suspension in the $y$ direction so that video of all parts of the suspension was recorded. The speed of scanning was $1$~mm/s. The particle observation cameras operated at a rate of $70$ frames per second.

\begin{table}[tbp]
\caption{Experimental parameters and estimate of the kinematic viscosity of complex plasma in four experimental runs.} \label{tab1}

\begin{tabular}{lcccc}
    \hline
    \hline
Parameter                                            & Run 1& Run 2& Run 3& Run 4           \\
    \hline
gas pressure $p_{\rm Ne}$ (Pa)                       & 40   & 60   & 60   & 15              \\
output laser power $P_{\rm laser}$ (W)               & 2.16 & 2.16 & 0.40 & 1.26            \\
electron density $n_e$ $(10^8$\,cm$^{-3})$           & 1.43 & 1.49 & 1.49 & 0.92$^a$        \\
electron temperature $T_e$ (eV)                      & 8.6  & 8.4  & 8.4  & 9.8$^a$         \\
screening length $\lambda_D$ ($\mu$m)                & 98   & 96   & 96   & 122$^a$         \\
particle diameter $2r_p$ ($\mu$m)                    & 3.38 & 3.38 & 3.38 & 6.86            \\
particle charge $Q$ ($e$)                            & 1900 & 2000 & 2000 & 5600$^a$        \\
particle number density$^b$ $n$ ($10^5$\,cm$^{-3})$  & 1.3  & 1.3  & 1.2  & 0.5             \\
Wigner-Seitz radius$^b$ $a$ ($\mu$m)                 & 122  & 123  & 127  & 173             \\
screening parameter $\kappa=a/\lambda_D$             & 1.2  & 1.3  & 1.3  & 1.4             \\
Epstein gas drag rate $\gamma$ (${\rm s}^{-1}$)      & 102  & 153  & 153  & 19              \\
kinematic viscosity $\nu_{\rm max}$ (${\rm mm^2/s}$) & 0.9  & 1.1  & 6.7  & 0.2             \\
normalised viscosity $\eta^*_{\rm max}$              & 0.29 & 0.33 & 1.95 & 0.05            \\
    \hline
    \hline
\end{tabular}
$^a$ For $p_{\rm Ne}=20$~Pa, $^b$ measured without shear flow.
\end{table}

Shear flow in the particle suspension was created by applying the radiation pressure force from the focused beam of a powerful manipulation laser. The laser beam had a diameter of $1.5$~mm (at a level of $1/e^2$) and was aligned with the discharge tube axis. The laser output power was in the range of $0.4$--$2.16$~W. This experiment was performed for various combinations of the experimental parameters (gas pressure, particle size, and laser power), see Table~\ref{tab1}.

A detailed quantitative analysis of the shear flow requires the precise knowledge of the manipulation-laser beam intensity profile at the position of particles. It is rather complicated and not known with sufficient accuracy \cite{Pustylnik:2017}. In the present experiments, we employed a new method of measuring the laser beam profile {\it in situ}, which is only possible in microgravity conditions: The plasma was briefly (during $0.3$--$0.5$~s) switched off while the manipulation laser was on. During the plasma off time, the particle charge rapidly declined \cite{Ivlev:2003} and the interparticle interactions all but vanished; however, the particle suspension did not collapse due to the absence of gravity. Instead, each particle attained terminal velocity (in the axial direction) due to the balance of the laser force and the neutral gas drag and the resulting particle velocity profile reproduced the laser beam intensity profile.

We used video recorded by camera 1 during a scan of a steady-state shear flow to reconstruct the particle flow field using the following method. In each frame, individual particles were identified using a moment method and then traced to the next frame. This gave the particle velocity components $v_{x,z}$. These 2D velocity fields were then stacked into a 3D flow field (taking into account the speed of scanning). From the obtained 3D flow field, various projections or cross sections can be calculated.

{\it Results.} The reconstructed longitudinal velocity profile $v_x(y,z)$ is shown in Fig.~\ref{vx_y_z}. (Scanning was performed in the $y$ direction.) This profile is valid under the assumption of stationary flow. The flow apparently has cylindrical symmetry. Therefore, it can be described by a flow profile $v_x(r)$, thus presenting a one-dimensional problem.

\begin{figure}[tb]
\centering
\includegraphics[width=0.8\columnwidth]{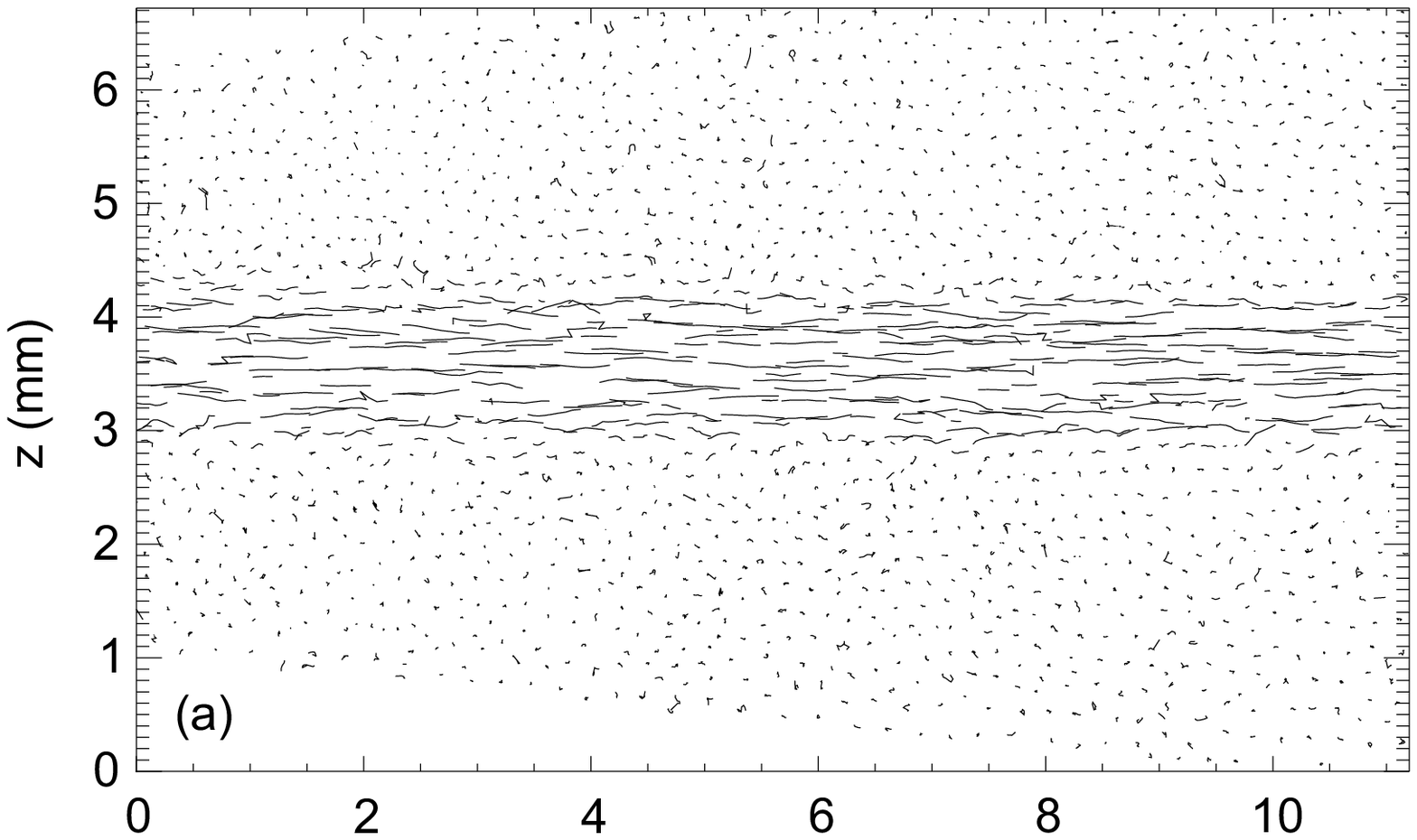}
\includegraphics[width=0.8\columnwidth]{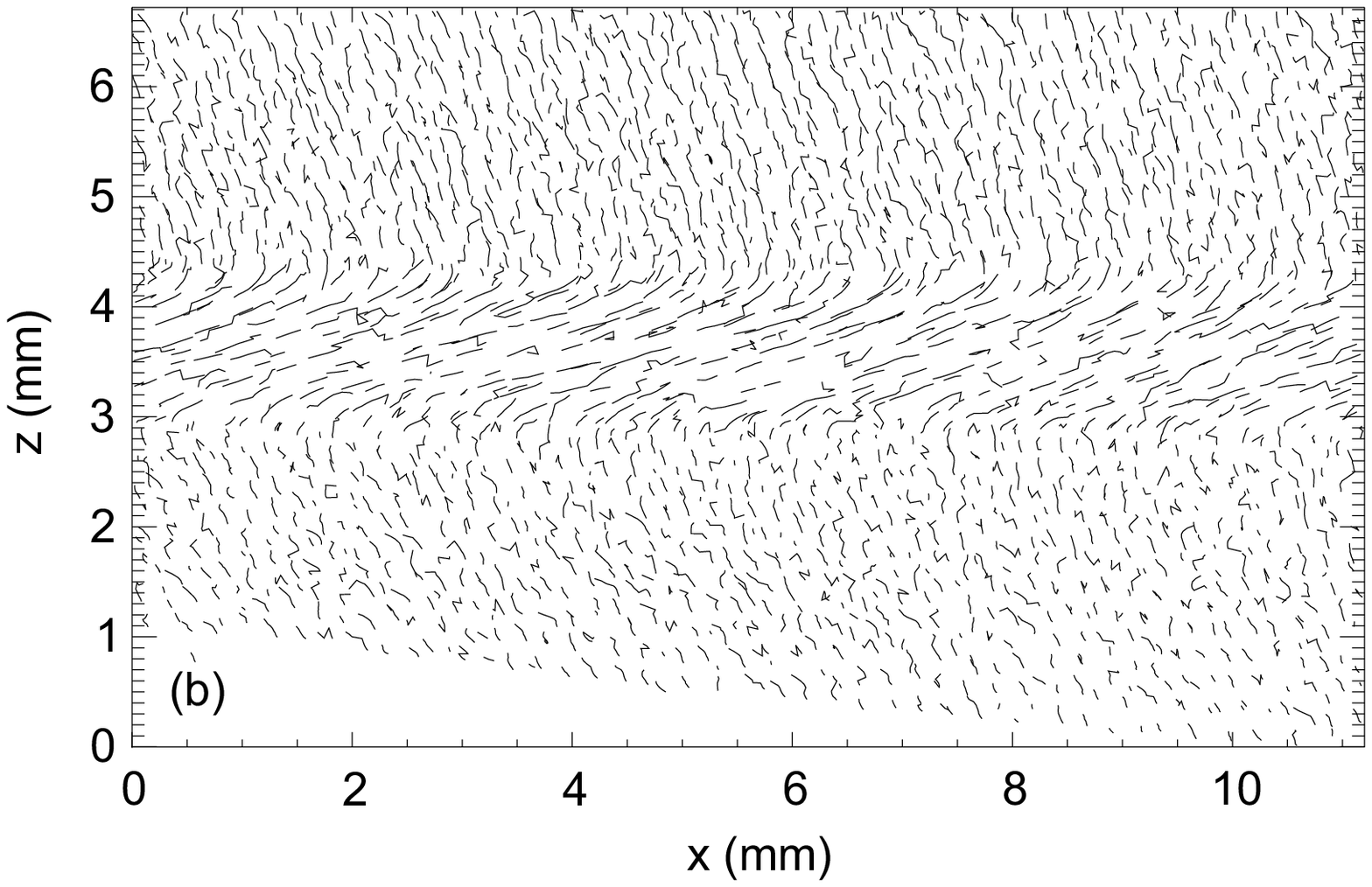}
\caption {\label {Trajectories} Particle trajectories during $0.143$~s measured in the central cross section of the shear flow in the (a) plasma on and (b) plasma off periods in experiment 1 in Table~\ref{tab1}. Corresponding particle velocity profiles $v_x(z)$ are shown in Fig.~\ref{vx_y}(a).
}
\end{figure}

Due to the cylindrical symmetry of the shear flow, it is sufficient to analyse its central cross section, as we do below. The particle trajectories during the plasma on and off periods in the experiment 1 in Table~\ref{tab1} are shown in Fig.~\ref{Trajectories}. Note that the actual trajectories are in general three dimensional and therefore may not be completely captured in these figures. The action of the manipulation laser is clearly seen in the middle of both panels where the particle trajectories are elongated in the $x$ direction.

\begin{figure}[tb]
\centering
\includegraphics[width=1.0\columnwidth]{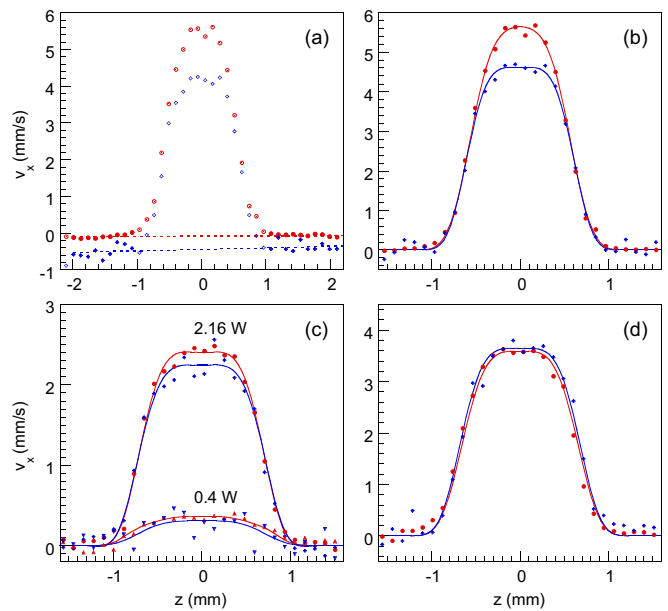}
\caption {\label {vx_y} Particle velocity profiles $v_x(z)$ (red circles, measured in the plasma on periods) and $v_x^{\rm off}(z)$ (blue diamonds, measured in the plasma off periods) in four experiments, see Table~\ref{tab1}: (a) and (b) experiment 1, (c) experiments 2 and 3, and (d) experiment 4. The origin in $z$ is shifted to the profile midpoint. In (a), the background velocity is shown by dashed lines (fits through solid symbols). In (b),(c),(d), the background velocity was subtracted from the flow profiles; the curves are fits with the empirical function $(a+bz^2+cz^6)\exp(-z^4/d^4)$, with $b=0$ for (c) 0.4 W.
}
\end{figure}

The time-averaged particle velocity profiles $v_x(z)$ and $v_x^{\rm off}(z)$ for the plasma on and off periods, respectively, are shown in Fig.~\ref{vx_y}. The background velocity was calculated by linear fits as shown in Fig.~\ref{vx_y}(a) and then subtracted from the velocity profiles as shown in Figs.~\ref{vx_y}(b),(c),(d). One can make two important observations (in the figures with subtracted background). First, $v_x(z) \gtrsim v_x^{\rm off}(z)$ for $|z|<0.5$~mm. While the reason for this is not completely clear, we note that the difference between $v_x(z)$ and $v_x^{\rm off}(z)$ is larger for higher laser power and may therefore be due to the particle heating. Second, $v_x(z) \simeq v_x^{\rm off}(z)$ for $|z|>0.5$~mm.

The latter result means that the shear viscosity of the particle suspension is low. More precisely, it means that $\nu / \gamma \equiv \ell_{\rm visc}^2$, where $\nu$ is the kinematic viscosity of the complex plasma liquid, $\gamma$ is the neutral gas drag rate, and $\ell_{\rm visc}$ is the momentum transport length, is not resolved here (within experimental error). In this situation, it is not possible to measure $\nu$, but at least it is possible to place an upper estimate on it, as we show below.

This conclusion is based on the assumption of fast and deep enough particle decharging during the plasma off period. Decharging of particles in a decaying afterglow plasma is a complex problem; for radio-frequency plasmas it was studied in detail in Ref.~\cite{Ivlev:2003}. It was reported that the afterglow plasma decayed within a few milliseconds after switching the discharge off, while the rest charge on the particles was around $1.6\times 10^{-2}Q_0\simeq 160e$, where $Q_0$ is the equilibrium charge of the particles in plasma.

In our experiments, we observed that when the discharge was switched off, the particle cloud initially did not undergo any dramatic change and its mean interparticle spacing [calculated from $g(r)$] increased by a maximum of $5$\%. The cloud started to slowly drift in the positive $z$ direction, see Fig.~\ref{Trajectories}(b), presumably due to the thermophoretic force caused by the inhomogeneous heating of the discharge tube, similar to the findings of Ref.~\cite{Ivlev:2003}. This means that the residual charge on the particles, if any, was small. Therefore, in the analysis below we neglect the interparticle interactions in the plasma off period. In some experiments, e.g., run 2 in Table~\ref{tab1}, the initial phase of slow drift was followed by the second ``fly-out'' phase: Approximately $0.17$~s after the discharge was switched off, the particles started to accelerate in the positive $x$ direction and spread in the $z$ direction. We do not consider the fly-out phase here.

\begin{figure}[tb]
\centering
\includegraphics[width=0.8\columnwidth]{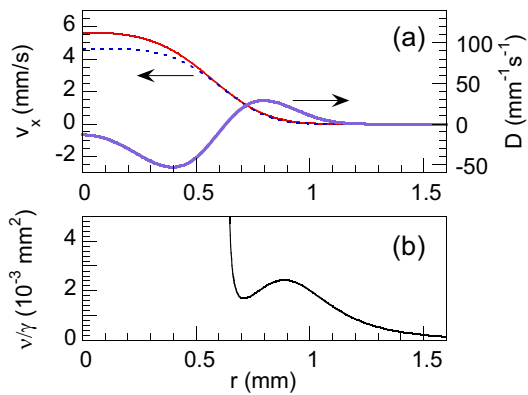}
\caption {\label {fits} (a) Fitted particle velocity profiles for the plasma on (thin red curve) and plasma off (dotted blue curve) periods and $\mathcal{D}$ (thick purple curve). (b) Plot of $\nu /\gamma$ for experiment 1 in Table~\ref{tab1}, see the text for details.
}
\end{figure}

To analyse the shear flow, we start with the Navier-Stokes equation (fluid equation where the particle suspension is treated as a continuous viscous liquid). For our situation (steady-state laminar flow of incompressible liquid with cylindrical symmetry) it reads:
\begin{equation}
\label{NStEq}
m_d\nu\frac{1}{r}\frac{d}{dr}(r\frac{dv}{dr})=m_d\gamma v-F_{\rm laser}(r),
\end{equation}
where $m_d$ is the dust particle mass, $\nu$ is the kinematic viscosity (assumed to be constant) of the complex plasma liquid, $v=v_x(r)$ is the flow velocity, $r$ is the radial coordinate, $\gamma$ is the neutral gas drag rate, and $F_{\rm laser}(r)$ is the laser force. Since $F_{\rm laser}=m_d\gamma v^{\rm off}$, where $v^{\rm off}=v_x^{\rm off}(r)$, and using the notation $\mathcal{D}=r^{-1}\frac{d}{dr}(r\frac{dv}{dr})$, Eq.~(1) can be written as $\nu\mathcal{D}=\gamma(v-v^{\rm off})$, or $(\nu/\gamma)\mathcal{D}=v-v^{\rm off}$. Here, $\mathcal{D}$ is well defined for the fitted smooth velocity profiles, see Fig.~\ref{fits}(a). However, since in the profile tails $v\simeq v^{\rm off}$ within the experimental error $\Delta v$ (defined as the rms deviation of measurements from the fitting curve), $\nu/\gamma$ is small and poorly defined. For example, the result of formally calculating $\nu/\gamma=(v-v^{\rm off})\mathcal{D}^{-1}$ for the data of Fig.~\ref{vx_y}(b) is shown in Fig.~\ref{fits}(b). In the tail of the flow velocity profile, $\nu/\gamma \simeq 10^{-3}~{\rm mm}^2$ and $\nu\simeq 0.1~{\rm mm^2/s}$. In some experimental runs, $v<v^{\rm off}$ in the fitted profile tails, which formally gives negative (unphysical) viscosity.

Further experiments were performed, where the experimental procedure was modified in a way that would increase the complex plasma's shear viscosity and render it measurable. Three approaches were tested. First, the manipulation laser power settings in the range from medium to very low (just above the lasing threshold) were used. The idea was to reduce shear thinning \cite{Nosenko:2013} so that $\nu$ would become larger and measurable. The outcome was qualitatively the same, i.e., $v\simeq v^{\rm off}$ within the experimental error, see Fig.~\ref{vx_y}(c). At the lowest laser power used ($0.4$~W), the particle velocity profiles are very noisy. Second, larger particles ($6.86~\mu$m in diameter) were used to achieve larger particle charge and therefore larger $\nu$ and simultaneously lower gas drag rate $\gamma$. Third, lower gas pressure ($15$~Pa) was used to achieve lower $\gamma$. These modifications of the experimental procedure were intended to increase the ratio $\nu / \gamma$ to a measurable level. The outcome of this test, however, was qualitatively the same, i.e., $v\simeq v^{\rm off}$ within the experimental error, see Fig.~\ref{vx_y}(d).

Therefore, it is not possible to reliably measure the complex plasma's shear viscosity in our experimental conditions, but it is possible to place an upper estimate on it. For example, in experiment 1 in Table~\ref{tab1}, $\nu < \nu_{\rm max} \equiv \gamma \, \Delta v \, \langle\mathcal{D}^{-1}\rangle \simeq 102\,{\rm s}^{-1} \times 0.093\,{\rm mm/s}\times 0.09\,{\rm mm\,s} \approx 0.9~{\rm mm^2/s}$, where the averaging is performed in the velocity profile tail $0.7~{\rm mm}\leq r\leq 1.1~{\rm mm}$. Similar estimates are also found for other experimental runs, see Table~\ref{tab1}.

{\it Comparison with previous results.} Surprisingly, our experimental results for kinematic viscosity turn out to be much lower than those reported in previous (ground-based) experiments with 3D complex plasmas: $\nu \approx 130~{\rm mm^2/s}$ in Ref.~\cite{Gavrikov:2005}, $\nu=0.8$--$160~{\rm mm^2/s}$ in Ref.~\cite{Vorona:2007}, and $\nu=10$--$300~{\rm mm^2/s}$ in Ref.~\cite{Ivlev:2007}. All these earlier experiments were designed in a fashion similar to ours: A shear flow in a complex plasma was created by laser manipulation. In Ref.~\cite{Ivlev:2007}, a prototypal laboratory PK-4 setup was used. Yet except for one measurement in Ref.~\cite{Vorona:2007}, the previously reported viscosity values are significantly higher than our estimate.

The physics behind this difference may be a different structure of the complex plasma in our experiments. In particular, it was shown that in microgravity conditions particles tend to form strings elongated along the discharge tube \cite{Pustylnik:2020}. String formation (which is a known mechanism of shear thinning in simulated colloids \cite{Rigos:1992}) can reduce the viscosity of complex plasma. Shear thinning and related shear-induced particle reordering were also experimentally observed in a laser-induced shear flow in a 2D complex plasma, where particles formed strings aligned in the flow direction \cite{Nosenko:2013}. It is interesting to note that our result for kinematic viscosity of a 3D complex plasma is comparable to that of a 2D complex plasma $\nu_{\rm 2D}=1.4$--$6~{\rm mm^2/s}$ \cite{Nosenko:2004,Nosenko:2013} and also of liquid water, $\nu_w\simeq1.8~{\rm mm^2/s}$ \cite {Morfill:04}.

Improving our estimate of the shear viscosity or even measuring its exact value will require extended experimental parameter ranges (first of all, much lower gas pressures) and therefore new flight hardware. This is not practical at the moment and must be reserved for future projects. Meanwhile, the results of the present work may help in designing a more advanced successor to the PK-4 instrument.

To compare our results with previous MD simulations, we calculate the normalised viscosity $\eta^*=\nu/(\omega_{\rm pd}a^2)$, where $\omega_{\rm pd}=(Q^2n/\epsilon_0m_d)^{1/2}$ is the dust plasma frequency. For example, in experiment 1 in Table~\ref{tab1}, $\omega_{\rm pd}=211~{\rm s}^{-1}$ and $a=0.122$~mm; therefore, $\eta^*_{\rm max}\simeq0.29$. For the conditions of experiment 1 ($\Gamma\simeq200$ and $\kappa=1.2$), the interpolation formula proposed in Ref.~\cite{Khrapak:2018} (based on equilibrium 3D MD simulations with Yukawa interparticle interactions) gives $\eta^*=0.22$. This is in agreement with our experimental estimate. The MD simulation of Ref.~\cite{Murillo:01} gives $\eta^*=0.06$ for the closest reported values of $\Gamma=100$ and $\kappa=1$. While also compatible with our experimental findings, this value is lower than that given in Ref.~\cite{Khrapak:2018}. This may be due to the nonequilibrium method used in Ref.~\cite{Murillo:01} (an imposed sinusoidal velocity profile was allowed to relax), which may model our experiment better than the equilibrium simulations based on the Green-Kubo formula. In this regard, we note the need of further computer simulations with a more realistic (anisotropic) interparticle potential \cite{Kompaneets_PhD} and using nonequilibrium methods.

{\it Acknowledgments.} The authors gratefully acknowledge the joint ESA-Roscosmos experiment ``Plasmakristall-4'' on board the International Space Station. A. D. U. and A. M. L. were supported by the Russian Science Foundation Grant No. 20-12-00365 and participated in preparation of this experiment and execution of it onboard the ISS. This work was supported in part by DLR/BMWi Grants No. 50WM1441 and No. 50WM1742. We thank Ch. Knapek for carefully reading the manuscript.

\end{document}